\documentclass[%
 reprint,
 amsmath,amssymb,
 aps,
]{revtex4-1}

\usepackage{graphicx}
\usepackage{dcolumn}
\usepackage{bm}

\begin{document}

\preprint{APS/123-QED}

\title[]{Transition from Dirac Points to Exceptional Points in Anisotropic Waveguides}

\author{Jordi Gomis-Bresco$^1$}

\author{David Artigas$^{1,2}$}
\email{david.artigas@icfo.eu}

\author{Lluis Torner$^{1,2}$}

\affiliation{1. ICFO-Institut de Ci\`{e}ncies Fot\`{o}niques, The Barcelona Institute of Science and Technology,  08860 Castelldefels (Barcelona),  Spain
}
\affiliation{2. Department of Signal Theory and Communications, Universitat Polit\`{e}cnica de Catalunya,  08034 Barcelona, Spain
}%

\date{\today}

\begin{abstract}
\noindent We uncover the existence of Dirac and exceptional points in waveguides made of anisotropic materials, and study the transition between them. Dirac points in the dispersion diagram appear at propagation directions where the matrix describing the eigenvalue problem for bound states splits into two blocks, sorting the eigenmodes either by polarization or by inner mode symmetry. Introducing a non-Hermitian channel via a suitable leakage mechanism causes the Dirac points to transform into exceptional points connected by a Fermi arc. The exceptional points arise as improper hybrid leaky states and, importantly, are found to occur always out of the anisotropy symmetry planes. \end{abstract}

\pacs{Valid PACS appear here}
\maketitle

\noindent Many physical phenomena that initially arose in quantum and solid-state physics and where rare bound states and special dispersion properties play a central role have found important and fertile implementations in optical systems. Chiral edge states \cite{haldane_possible_2008, wang_observation_2009}, Weyl points \cite{lu_weyl_2013}, topological insulators \cite{rechtsman_strain-induced_2013,khanikaev_photonic_2013, hafezi_imaging_2013,parto2018edge,bandres2018topological,harari2018topological}, or bound states in the continuum \cite{zhen_topological_2014, doeleman_experimental_2018, gomis-bresco_anisotropy-induced_2017}, to cite only a few, are examples of effects  
that have opened rich lines of research that are of continuously growing interest for both, the fundamental understanding of wave phenomena and its application to photonic devices. Occurrence of Dirac points (DPs) and exceptional points (EPs) are another salient example.

By and large, Dirac points are singularities in the band diagrams of Hermitian systems that are at the core of the unique properties of the corresponding structures and materials, as for example in the electronic properties of graphene \cite{castro_neto_electronic_2009}. A DP occurs when two bands cross each other locally and exhibit a linear dispersion in any direction in the momentum space \cite{haldane_possible_2008}. As the eigenvalues of Hermitian systems are real, two orthogonal eigenstates coexist at the DP with the same eigenvalue.  The counterpart in non-Hermitian systems are exceptional points \cite{feng_non-hermitian_2017}, where the complex eigenvalues of two different bands are identical, with equal real and imaginary parts. In EPs, the eigenvectors and therefore the bands are also degenerate. Thus, at an EP the matrix describing the system in standard formalism as an eigenvalue problem cannot be diagonalized. Such properties result in unique dynamics near EPs \cite{miri2019exceptional}, which result in, e.g., asymmetric mode switching \cite{doppler_dynamically_2016,yoon_time-asymmetric_2018}, appearance of polarization topological half-charges \cite{zhou_observation_2018}, chiral modes and directional lasing \cite{peng2016chiral}, or ultrasensitive measurements \cite{chen2017exceptional, hodaei2017enhanced}. 

Adding a non-Hermitian physical effect transforms DPs into EPs \cite{ozdemir_fermi_2018}. Studying the transition between them requires a system where DPs exist and EPs can be generated by opening a non-conservative channel. In this Letter we address the existence conditions of DPs in waveguiding structures containing uniaxial anisotropic materials and study their transformation into EPs when a tunable leakage mechanism opens a radiation channel. Encircling the EPs can be conceptually done by varying the optical axis orientation relative to the propagation direction of the material forming the waveguide core.

\begin{figure}
\includegraphics[width=0.45\textwidth]{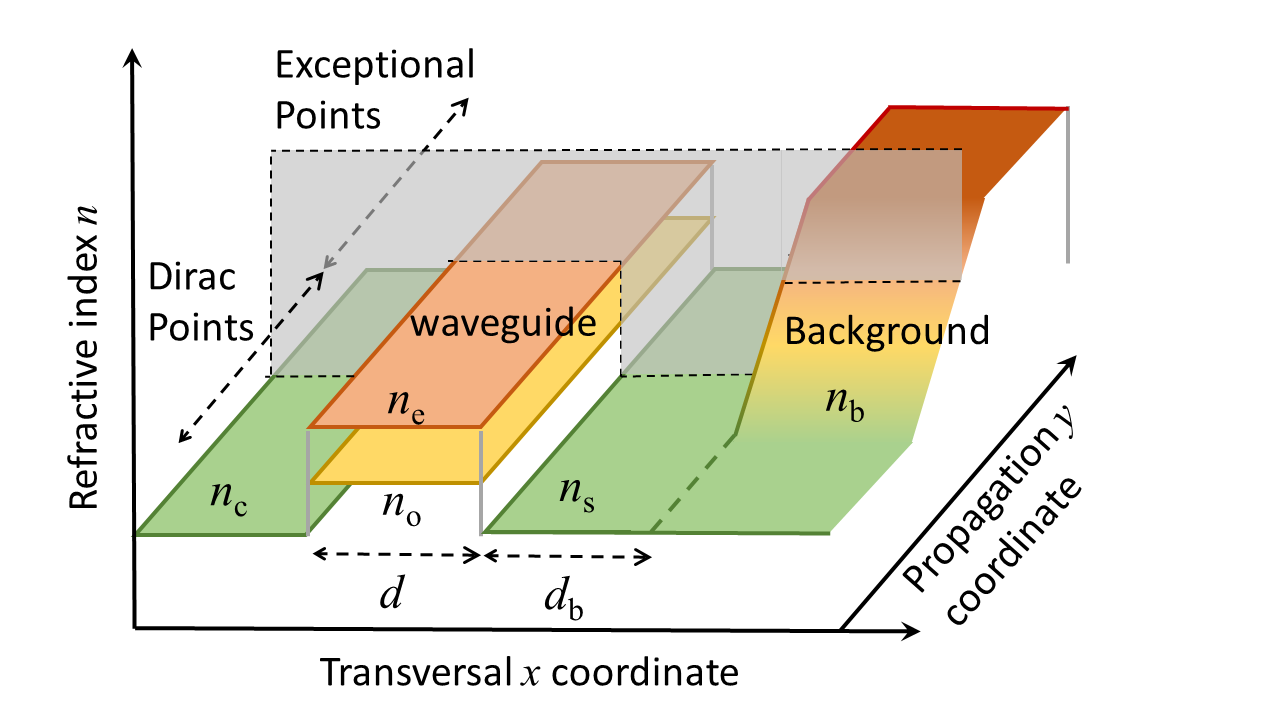}
\caption{\small Schematic minimal waveguiding structure comprising an isotropic cladding and a substrate with refractive indices $n_c$ and $n_s$, respectively, and a guiding film made of an uniaxial anisotropic medium with thickness $d$ and ordinary and extraordinary refractive indices $n_o$ and $n_e$. In the second half of the structure, a tunable refractive index $n_b$, which may be induced externally, e.g., by a thermooptic effect, is located at a distance $d_b$ from the guiding film. The grey plane separates the regions of Hermitian and non-Hermitian behavior. \label{f1}}
\end{figure}

\begin{figure}
\includegraphics[width=0.45\textwidth]{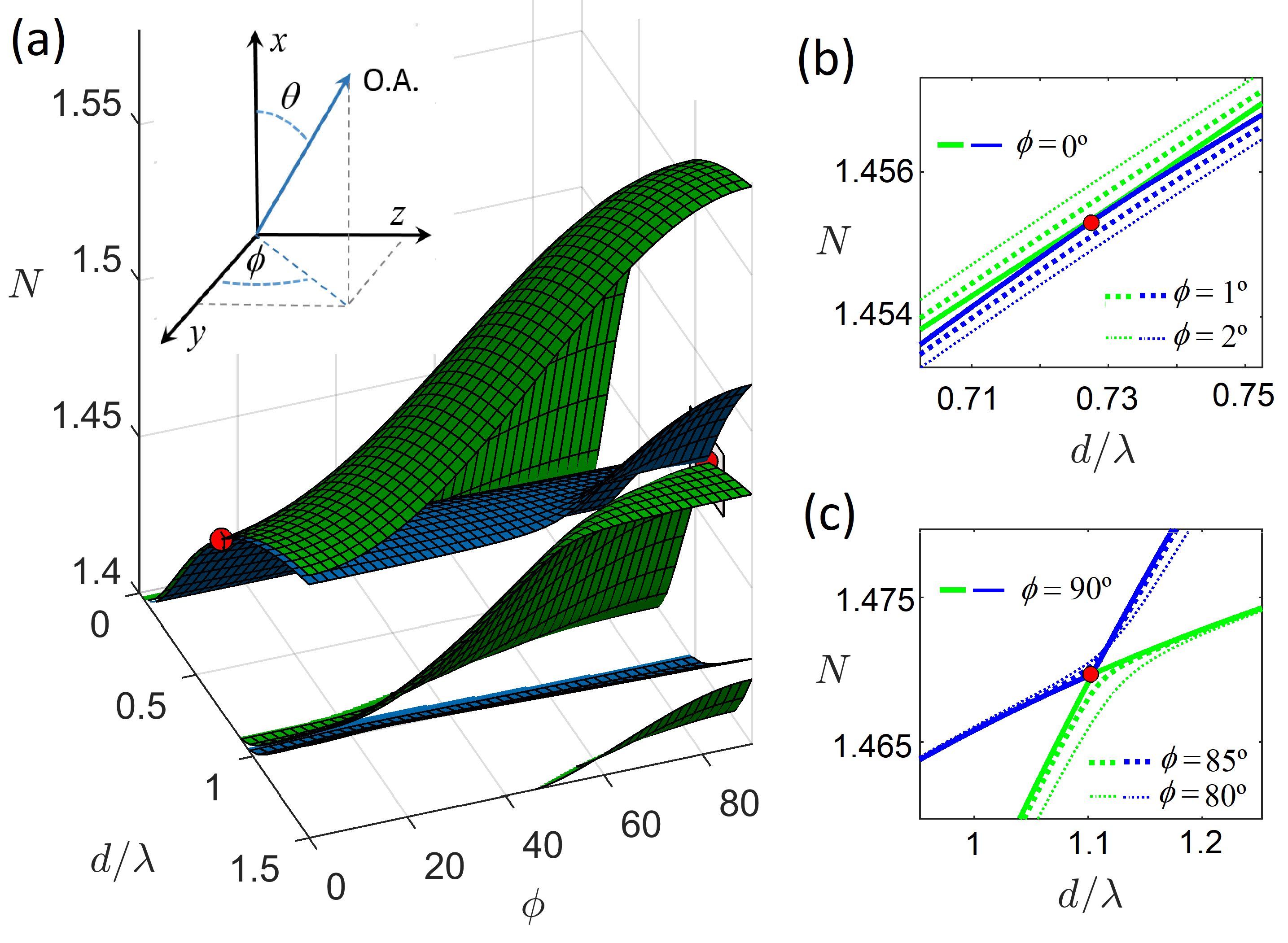}
\caption{\small (a) Dispersion bands of the eigenmode effective index $N$ depicted in green and blue colors, as a function of the normalized film thickness $d/\lambda$ and propagation direction, $\phi$. The arrow labelled OA in the inset indicates the optical axis orientation with polar $\theta$ and azimuth $\phi$ angles in spherical coordinates. The waveguide parameters are $n_c=n_s=n_b=1.4, n_e=1.6, n_o=1.5$ and $\theta=80^{\circ}$. The dispersion diagram can be transformed into energy $E$ vs.\ momentum $k_y$-$k_z$ diagrams by  setting $E=\hbar\omega\propto d/\lambda$,  $k_y=Nk_0\cos{\phi}$ and $k_z=Nk_0\sin{\phi}$. Two DPs are indicated as red dots at the crossings between the fundamental eigenmodes existing at $\phi=0^{\circ}$, and between the fist even and second odd eigenmodes existing at $\phi=90^{\circ}$. The first DPs are TE/TM polarized states, while the second DPs are fully hybrid states. The zooms blow-up the areas near (b) $\phi=0^{\circ}$ and (c) $\phi=90^{\circ}$. DPs occur at the symmetry planes $\phi=0^{\circ}$ in (b) and $\phi=90^{\circ}$ in (c) while any other propagation direction shows anti-crossings. The structure and the dispersion diagram are symmetric with respect to the $\theta = 90^{\circ}$, $\phi=0^{\circ}$ and $\phi=90^{\circ}$ planes. \label{f2}}
\end{figure}

The existence of DPs in waveguiding structures can be elucidated by analysing the matrix describing the eigenvalue problem for bound states. Waveguides with isotropic materials are described by two independent matrices for Transverse Electric (TE) and Magnetic (TM) eigenmodes. This results in lines in the dispersion diagram that do not cross each other, therefore DPs do not exist. In contrast, general structures made of anisotropic materials are described by a matrix that cannot be separated in smaller parts, resulting in the intrinsic hybrid polarization of the eigenmodes. Solving the eigenvalue problem as a function of the propagation direction results into eigenmodes that exist in surfaces (bands) in the three dimensional dispersion diagram [Fig.\ref{f2}(a)]. However, under suitable material or geometrical symmetric conditions, waveguides made of anisotropic media also allow splitting the matrix into two blocks after suitable algebraic manipulations. Under such conditions, the resulting matrix provides also sets of eigenmodes described by two different eigenequations. At such propagation directions and at a given wavelength the corresponding  bands can cross each other and exhibit linear dispersion, therefore resulting in DPs.

We found that DPs exist for different planar waveguide parameters and anisotropy configurations. Their inner nature is best exposed by analysing a {\it symmetric\/} structure with a film made of an uniaxial crystal surrounded by isotropic materials, as in Fig.\ref{f1} with $n_b=n_s=n_c$. Wave propagation is set along the $y$ direction and the optical axis of the film is oriented at a direction given by the polar $\theta$ and azimuth $\phi$ angles. Calculations were performed using the formalism described in Ref.\ \cite{mukherjee_topological_2018} and elaborated in detail in the Appendix for the case addressed here. In Fig.\ref{f2} we plot the dispersion diagram for a structure with a film with positive birefringence and optical axis orientation pointing out-of-plane at $\theta=80^{\circ}$, plotted as the variation of the effective index $N=k/k_0$ (where $k$ is the wave momentum along the propagation direction and $k_0$ the vacuum wavenumber) versus the normalized film thickness $d/\lambda$ and the propagation direction, which by simple rotation is given by the value of the angle $\phi$. Two existing DPs are shown as red dots in Fig.\ref{f2}(a): the first one at $\phi=0^{\circ}$ occurs between the first two bands when the eigenmodes are TE- and TM-polarized. Fig.\ref{f2}(b) shows the DPs as a crossings at $\phi=0$, and anti-crossings for $\phi \neq 0$.  In an important physical insight, the second DP ($\phi=90^{\circ}$, Fig.\ref{f2}a) arises as a crossing between the second and third bands, at a propagation direction where the system matrix splits into two blocks, now describing {\it even\/} and {\it odd\/} eigenmodes instead. Importantly, note that in this last case  the modes at the DP are fully hybrid and that their existence is a phenomenon that occurs owing to the perfect symmetry of the structure; in asymmetric geometries such DPs cease to exist. We found that other DPs (not shown) appear between alternating bands when $d/\lambda$ is increased further. We also found that when the guiding film features a negative birefringence, the first DP appears for even and odd modes between the two first bands at $\phi = 90^{\circ}$, while the DP that exist at $\phi = 0^{\circ}$ between TE and TM modes in this case arises between the second and third bands. In all cases, {\it anisotropy is necessary for the DPs to occur\/}.

Mode crossings in waveguides made of anisotropic media are known to exist, see e.g., \cite{knoesen_hybrid_1988,yakovlev_fundamental_2003,satomura1974analysis,maldonado1996hybrid}. However, it must be properly appreciated that to date none of such crossings have been identified as a DP and, more importantly, {\it most crossings are not DPs\/}, actually. For example, the matrix describing the waveguide studied in Fig.~\ref{f2} splits into two blocks for two other configurations: when the optical axis is oriented orthogonal ($\theta=0^{\circ}$) and parallel ($\theta=90^{\circ}$) to the structure interface, resulting into pure TE/TM and even/odd eigenmodes, respectively \cite{marcuse_modes_1979}. However, none of such cases correspond to a DP, because the splitting of the matrix, and therefore the surface crossing, appear for any propagation direction within the waveguide plane, and not at specific directions, thus failing to show the linear dispersion dependence required for a DP.

Dirac points transform into EPs by introducing gain or losses in the system  \cite{zhen_spawning_2015,zhou_observation_2018,ozdemir_fermi_2018}. Thus, we open a non-conservative channel by placing a region with a high refractive index close to the film, which causes energy to leak away. The high refractive index may be provided by a suitable bulk material or may be induced externally, e.g., by a thermooptic effect. The loss strength is dictated by the penetration of the evanescent tails into the high-index material and, importantly, also by the  optical axis orientation, which affects the hybrid composition of the eigenmodes in terms of ordinary and extraordinary waves and thus the fraction of total energy carried by the component that becomes leaky. In the analysis, we set the refractive indices, the wavelength and the thickness of the waveguide, and vary the optical axis orientation.

\begin{figure}
\includegraphics[width=0.45\textwidth]{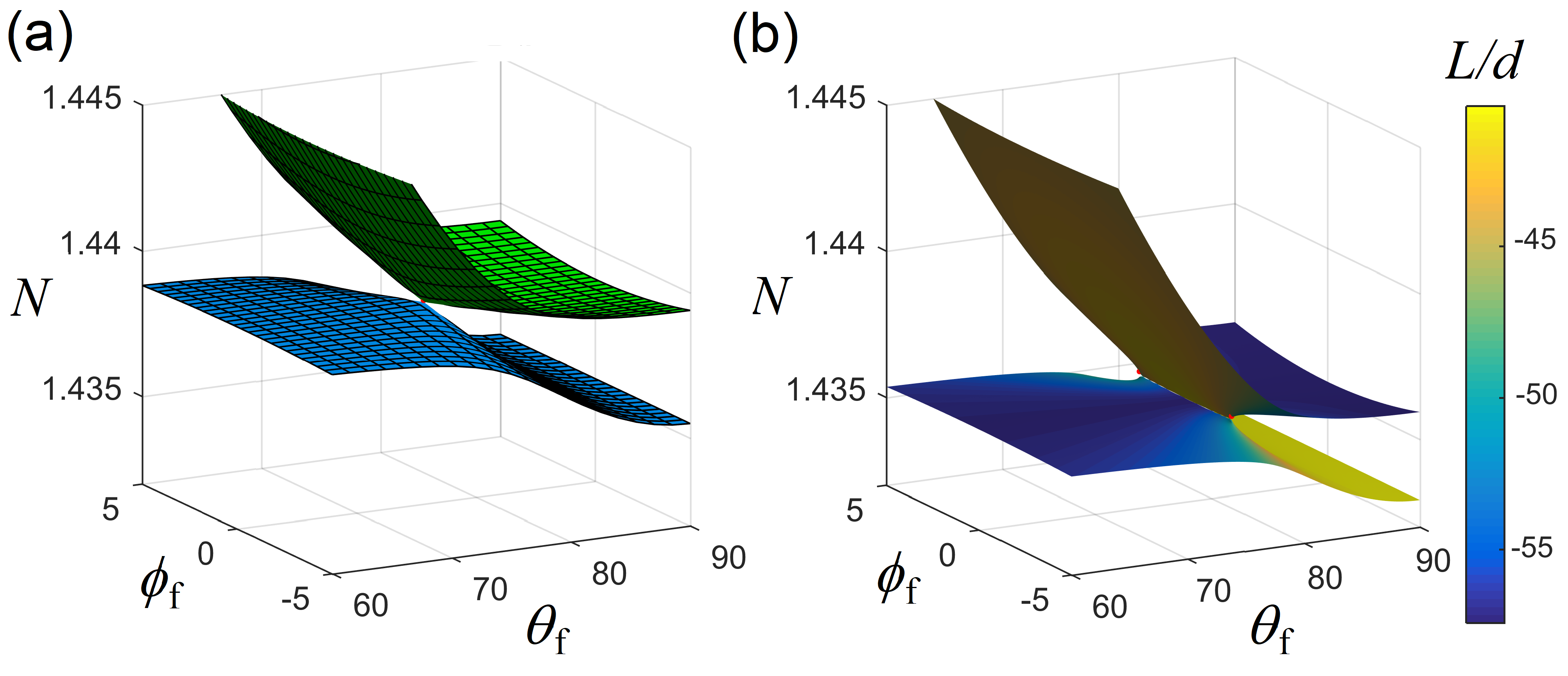}
\caption{\small (a) Dispersion for the waveguide in Fig. \ref{f2} as a function of $\theta$ and $\phi$  with  $d/\lambda=0.5$. (b) Same as in (a) but now for a structure with $n_b=1.8$ located at a distance from the film $d_b/\lambda=0.5$. The colour scale in (b) is proportional to the normalized decay length. \label{f3}}
\end{figure}

\begin{figure}
\includegraphics[width=0.42\textwidth]{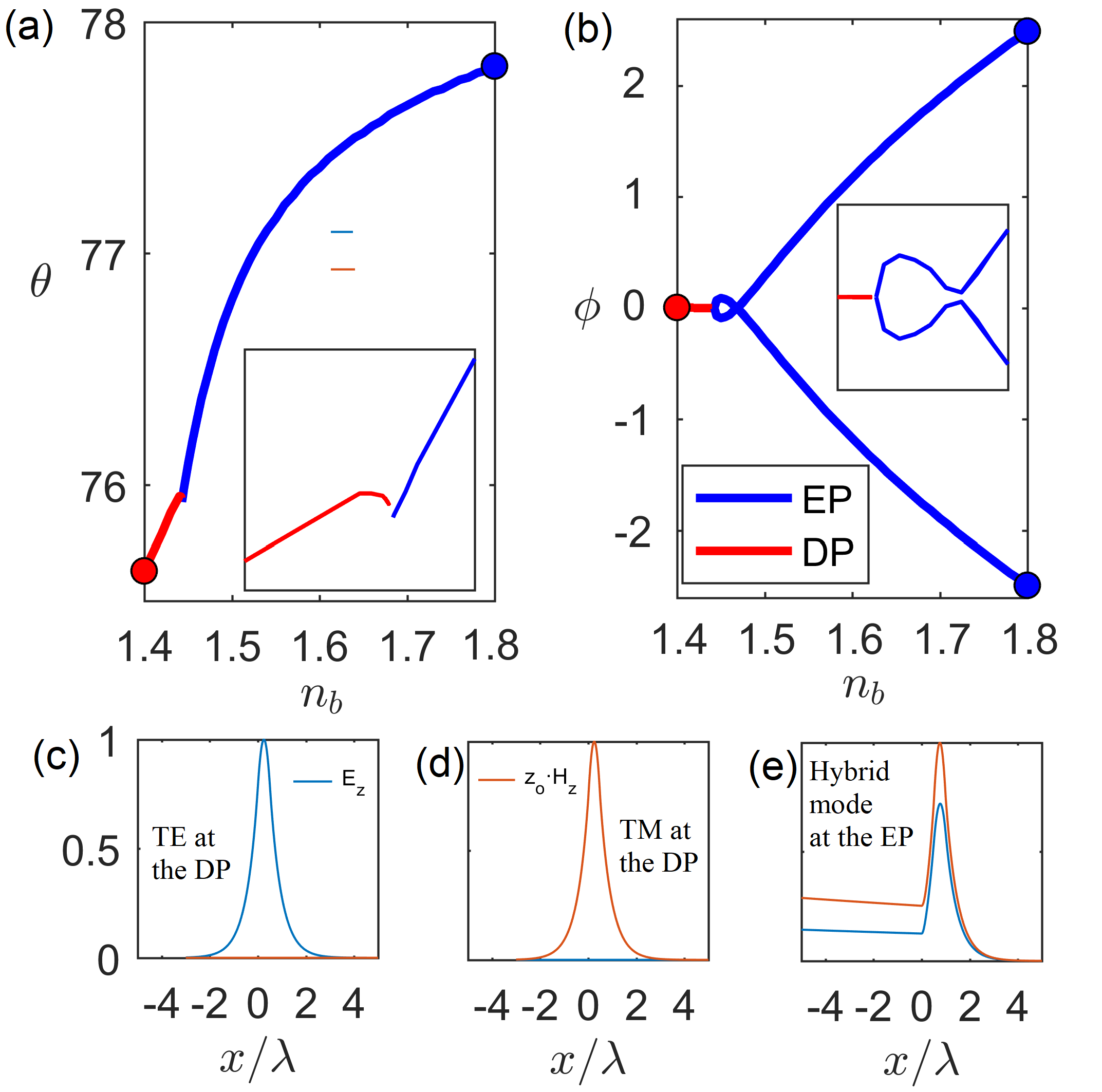}
\caption{\small (a), (b) Angular loci of the DPs (red) and EPs (blue) as a function of $n_b$ for $d_b/\lambda=0.5$. The insets expand the transition regions; the apparent gap in the inset of (a) is due to the finite resolution of the calculations. Red and blue dots correspond to  Fig.\ref{f3}(a) and \ref{f3}(b), respectively. Eigenmode of the (c) TE and (d) TM bound states at the DP. (e) Field components of the hybrid leaky mode near the EP. \label{f4}}
\end{figure}

Figure~\ref{f3}(a) shows the two bands existing above cutoff in the structure corresponding to Fig.~\ref{f2} for $d/\lambda=0.5$. The bands touch each other in a DP located at the optical axis orientation $(\theta=75.6^{\circ}, \phi=0^{\circ})$, where the eigenmodes are separable by polarization. Rising the  refractive index from $n_b=n_s$ to $n_b < N$ only changes the optical axis polar orientation $\theta$ at which the DP exists [Fig.~\ref{f4}(a)]. Yet, the system remains Hermitian, the DP is found at the symmetry plane $\phi=0^{\circ}$, and the polarization remains either TE [Fig.~\ref{f4}(c)] or TM [Fig.~\ref{f4}(d)]. In contrast, when $n_b > N$ a radiation channel is opened and the system becomes non-Hermitian. Then, the eigenmodes become {\it improper hybrid leaky modes\/} with complex $N$, and the DP transforms into a pair of EPs that, therefore, occur {\it out of the anisotropy symmetry planes\/}.

For slightly larger values of $n_b$, the two EPs occur closer to each other (see the inset) and, as $n_b$ keeps increasing, the loci at which the EPs are located depart further from the symmetry plane $\phi\neq 0^{\circ}$ [Fig.~\ref{f4}(b)]. A representative shape of the dispersion diagram of the leaky modes for $n_b=1.8$, featuring two EPs located at $\theta=77.78^{\circ}$, $\phi= \pm 2.49^{\circ}$, is shown in Fig.~\ref{f3}(b). At the EPs the bands coalesce, with $N$ having identical real and imaginary parts. The real part of the effective index of the leaky modes is identical at both bands in the line connecting the two EPs, a property that is equivalent to a Fermi arc in the energy-momentum dispersion diagram. The imaginary part of $N$ differs along the Fermi arc for the two bands, except at the EP where the modes are completely degenerate. These EPs are hybrid states [Fig.~\ref{f4}(e)] and are located at directions where the optical axis is oriented out of any anisotropy symmetry plane of the structure. We found that the TE/TM projections of the hybrid modes around the EPs remains almost constant, with the TM fraction being larger (almost twice in the particular case shown) than the TE polarization. 
\begin{figure}
\includegraphics[width=0.45\textwidth]{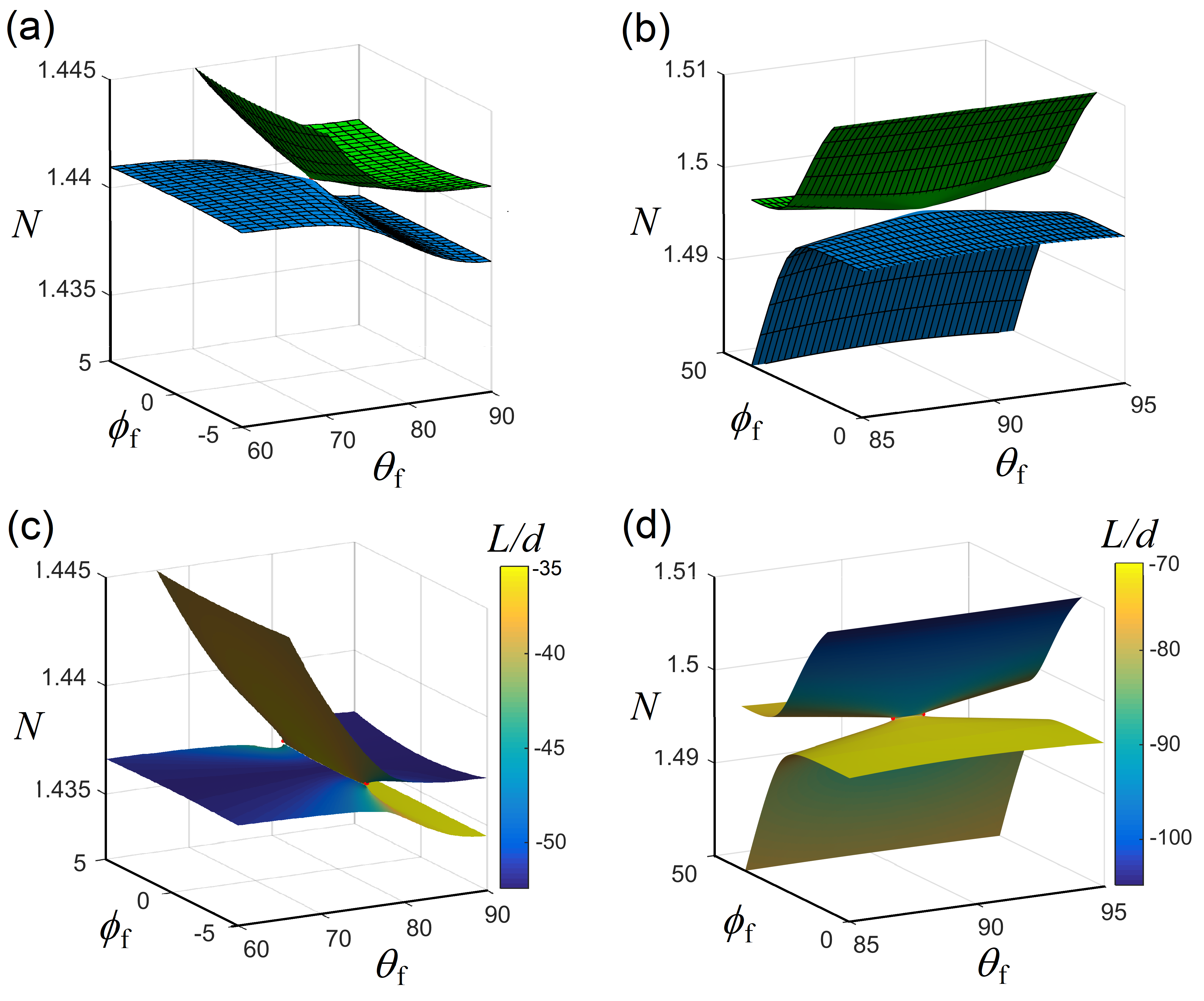}
\caption{\small (a) Dispersion of an asymmetric waveguide as a function of $\theta$ and $\phi$, for a system similar to the one considered in Fig.~\ref{f3}(a), but now $n_s=n_b=1.41$. (b) Same as in (a) but for a waveguide with a negative uniaxial film with various parameters: $n_c=1.4, n_s=n_b=1.41$, $n_e=1.5$, $n_o=1.6$, and $d/\lambda=0.5$. (c) and (d) correspond to the waveguides in (a) and (b), respectively, but coupled to an isotropic background with refractive index $n_b=1.8$ separated a distance 0.5 $d/\lambda$ from the film. The colour scales in (c) and (d) are proportional to the normalized decay length $L/\lambda$. \label{f5}}
\end{figure}

\begin{figure}
\includegraphics[width=0.4\textwidth]{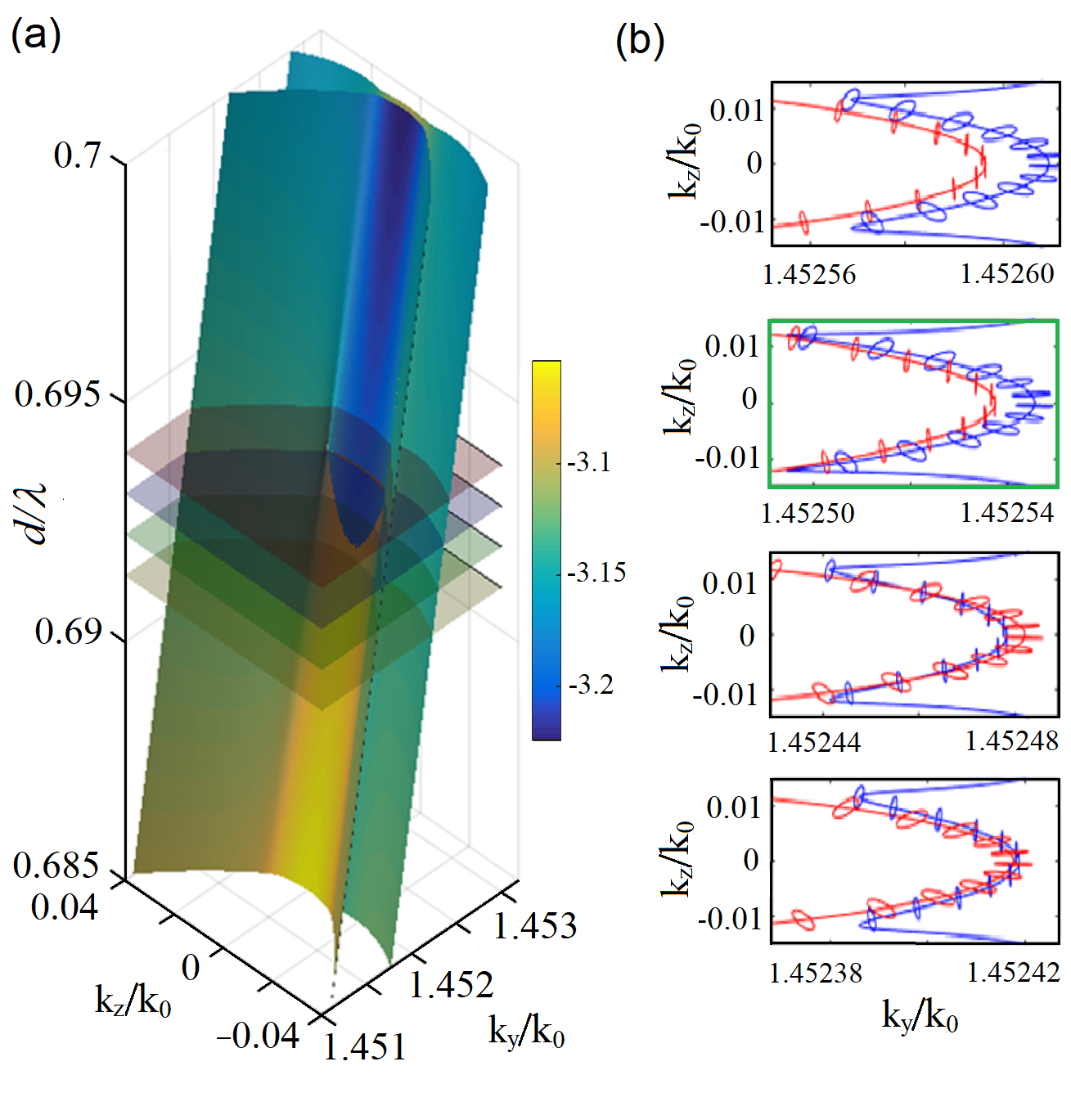}
\caption{\small Topological featuers of the radiating fields.  (a) Frequency-momentum dispersion diagram showing  EPs and a Fermi arc for the waveguide with isotropic background shown analyzed in Fig.~\ref{f5}(c). The colour scale is proportional to the normalized decay length $L/\lambda$. (b) Isofrequency cuts represented with the grey surfaces in (a) at which we plot the polarization of the radiation field. The second panel with green frame corresponds to an isofrequency that contains the EPs. \label{f6}}
\end{figure}

The mechanism that splits the matrix that describes the system impacts the robustness of the DPs against perturbations. DPs arising between bands with different polarization are robust even against asymmetric perturbations, as the system matrix can be split in any case. This yields the dispersion diagram shown in Fig.~\ref{f5}(a), which corresponds to a waveguide that is asymmetric in terms of the refractive index. However, DPs arising between bands with different parity cease to exist in the presence of asymmetric perturbations, as in such a case the system matrix can only be divided into blocks when the waveguide is symmetric, as elaborated in the  Appendix. As a consequence, a gap opens in the dispersion band diagram of asymmetric structures [Fig.~\ref{f5}(b)]. In contrast, EPs are robust again perturbations and  do appear in asymmetric non-Hermitian structures. Fig.~\ref{f5}(c) shows the two EPs related to the DP arising between bands with different polarization when the structure in \ref{f5}(a) is coupled to an isotropic background. In contrast to simple expectations, Fig.~\ref{f5}(d) also shows two EPs, even when a gap was present between the two bands with different parity in Fig.~\ref{f5}(b). In this case the origin of EPs must be seek in the DPs arising in symmetric structure rather than in the asymmetric one.  

Note that robustness of EPs has been related to the topological properties exhibited by the corresponding radiated fields \cite{zhou_observation_2018}. Fig.~\ref{f6} shows the polarization of the radiated field for the structure analyzed in Fig.~\ref{f5}(c), for different isofrequency cuts in the dispersion diagram. The isofrequency surface that encircles the EPs (the panel with green frame in Fig.~\ref{f6}(b)) shows the half-charge polarization winding near the EP described in \cite{zhou_observation_2018}. A half topological charge is apparent starting from the vertical red polarization, traversing the full contour in the clockwise direction and returning to the same point. Then the polarization flips direction by rotating $180^{\circ}$ in the clockwise direction.

\begin{figure*}
\includegraphics[width=0.8 \textwidth]{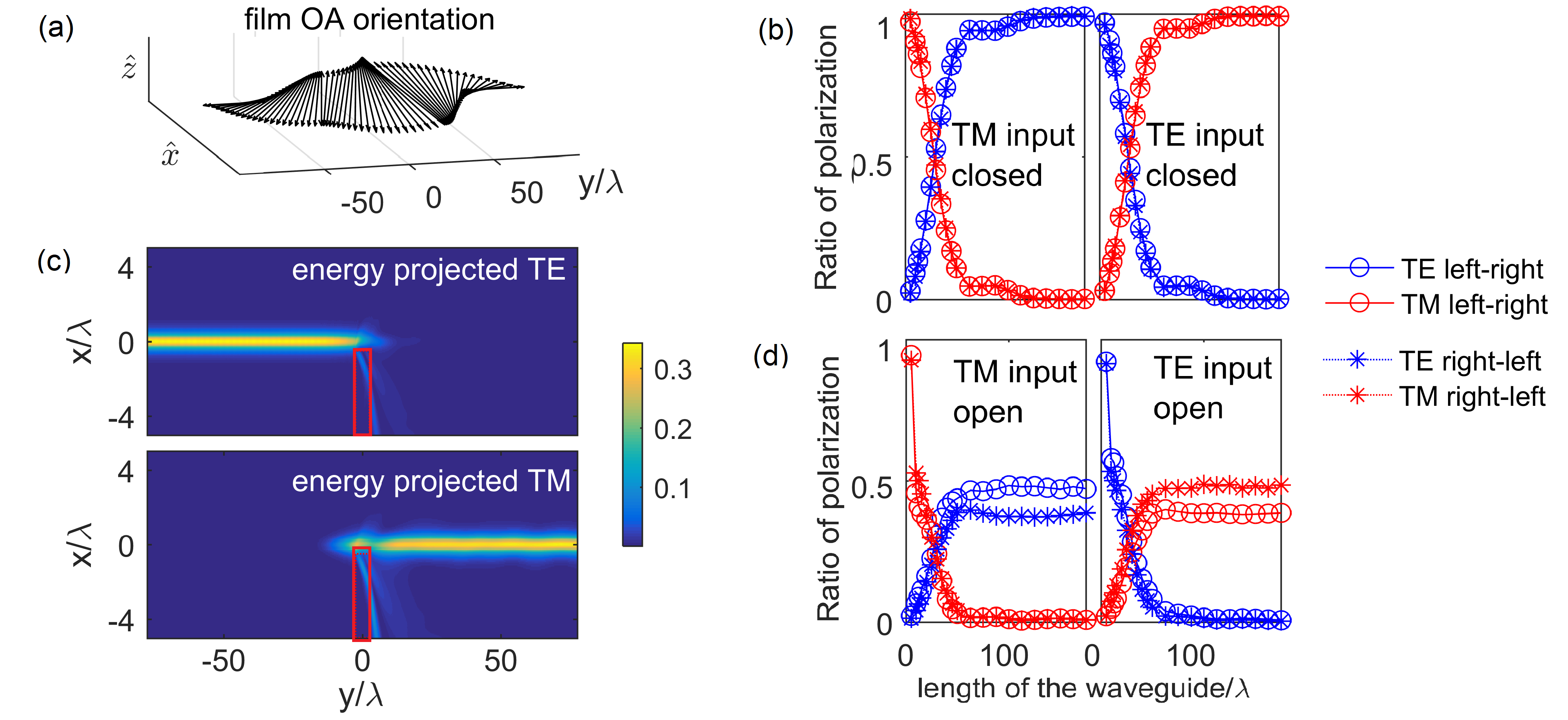}
\caption{\small Hermitian vs non-Hermitian dynamical evolution modelled with FDTD. (a) Variation of the film optical axis orientation along the $y$ axis used in all the calculations: The optical axis follows a closed circuit in the $\theta$ - $\phi$ parameter space. The calculations were performed for the waveguide characteristics as in Fig.\ \ref{f3}. (b) Mode conversion for the Hermitian structure as a function of the device length when a TM (left panel) and TE (right panel) mode is injected from the left (circles) and the right (asterisk) side of the structure. (c) Example of FDTD propagation combining Hermitian and non-Hermitian (red square) sections in the same structure. (d) Same as in (b) but when the central part of the structure is non-Hermitian. \label{f7}}
\end{figure*}

The transition from a Hermitian to a non-Hermitian behavior allows comparing the dynamical evolution in the proximity of DPs and EPs. In the anisotropic waveguides, this can be done by varying the optical axis orientation in the film along the propagation direction [Fig. \ref{f7}(a)]. We performed Finite-Difference-Time-Domain (FDTD) calculations \cite{miip} along a closed circuit in the $\theta - \phi$ parameter space in a clock- and anti-clockwise direction, which is equivalent to excite the structure from the right or left sides, and study reversal (direction-independent) versus chiral (direction-dependent) mode conversion \cite{doppler_dynamically_2016, hassan2017chiral, hassan_dynamically_2017, hassani_gangaraj_topological_2018}. In the Hermitian structure, the linear dispersion at the DP allows exchanging the band while maintaining the polarization, and the anti-crossing existing in their proximity (Fig.\ref{f2}(b,c)] results in polarization conversion \cite{torner-guided_1993}. We therefore chose an arbitrary (i.e., by no means optimized) closed circuit that crosses trough the DP and returns through an anti-crossing. A direct consequence of the election of a circuit that crosses a DP is conversion from a TE (TM) input to a TM (TE) output after returning at the initial point [Fig. \ref{f7}(b)]. The conversion is total and independent of the direction of excitation when the change in optical axis orientation is adiabatic, which in the case of the figure occurs for propagation lengths larger that $100\lambda$. In the case of a circuit containing a DP between bands with different parity [Fig. \ref{f2}(c)], the conversion refers to the parity state. 

When $n_b$ increases, the circuit encircles the EP in a non-Hermitian way. The change of polarization still occurs, as the branches are exchanged through the Fermi arc.  However, the conversion now is chiral and depends on the propagation direction.  The phenomenon occurs also with low losses, when the leakage mechanism is introduced only in one section of the circuit [the red square in Fig. \ref{f7}(c)]. Chiral conversion is illustrated in Fig. \ref{f7}(d), where the conversion from TE (TM) input to TM (TE) output yields a different value when the waveguide is excited from the right or from the left side. In this particular case the non-Hermitian section is short and it is located at the center of the waveguide, which is enough to expose the occurrence of the chiral behavior. However, its absolute strength may be enhanced by optimizing the location and length of the section containing the radiative channel. 

To summarize, we stress that the new feature introduced in this Letter is the existence of DPs and EPs in waveguiding structures made of anisotropic materials. The anisotropy introduces intrinsic angular-dependent propagation properties, which allows the existence of EPs  {\it out of symmetry planes\/}, and affords the possibility to follow the transition between DPs and EPs by opening and closing a suitable non-conservative channel. The transition from the Hermitian to a non-Hermitian behavior occurs via a radiation channel that generates leaky modes that are hybrid (with the full field components) and improper (they are modes of infinite energy that capture the physics of a infinite band of proper modes belonging to the continuum spectrum), which break the Hermitian behavior through radiation rather than via material losses. Such a physical mechanism is  general and should be applicable to other types of photonic structures showing similar phenomena, as is the case of photonic crystals with a graphene-like lattice \cite{Huahu2015, collins2016integrated, ma2016all, dong2017valley}. We studied a simple structure made of film birefringent materials, but the concept holds for more complex structures, including waveguides made of biaxial materials, of multiple anisotropic layers that allow a higher control of radiation, and of general anisotropic metamaterials.

\begin{acknowledgments}
\noindent
{\small Authors acknowledge financial support of the Generalitat de Catalunya through AGAUR 2017-SGR-1400 and CERCA; the Ministry of Science and Universities of Spain through FIS2015-71559-P and Severo Ochoa SEV-2015-0522; Fundaci\'o Cellex and Fundaci\'o Mir-Puig.}
\end{acknowledgments}

\section*{Appendix: Splitting of the system matrix}

\noindent The dielectric tensor of a uniaxial material for an arbitrary orientation of the optical axis given by the angles $\theta$ and $\phi$ as defined in Fig. 2(a) of the main text reads as
\medskip
\begin{widetext}
\begin{equation} 
\epsilon = \left[\begin{matrix}\epsilon_{xx} & \epsilon_{xy} & \epsilon_{xz}\\ \epsilon_{xy} & \epsilon_{yy} & \epsilon_{yz}\\ \epsilon_{xz} & \epsilon_{yz} & \epsilon_{zz}\end{matrix}\right] =
\left[\begin{matrix}\epsilon_{e} \cos^{2}{\left (\theta \right )} + \epsilon_{o} \sin^{2}{\left (\theta \right )} & \Delta \sin{\left (\theta \right )} \cos{\left (\phi \right )} \cos{\left (\theta \right )} & \Delta \sin{\left (\phi \right )} \sin{\left (\theta \right )} \cos{\left (\theta \right )}\\\Delta \sin{\left (\theta \right )} \cos{\left (\phi \right )} \cos{\left (\theta \right )} & \Delta \sin^{2}{\left (\theta \right )} \cos^{2}{\left (\phi \right )} + \epsilon_{o} & \Delta \sin{\left (\phi \right )} \sin^{2}{\left (\theta \right )} \cos{\left (\phi \right )}\\\Delta \sin{\left (\phi \right )} \sin{\left (\theta \right )} \cos{\left (\theta \right )} & \Delta \sin{\left (\phi \right )} \sin^{2}{\left (\theta \right )} \cos{\left (\phi \right )} & \Delta \sin^{2}{\left (\phi \right )} \sin^{2}{\left (\theta \right )} + \epsilon_{o}\end{matrix}\right],
\end{equation}
\end{widetext}
\medskip
where $\epsilon_o=n_o^{2}$ and $\epsilon_e=n_e^{2}$ are the ordinary and the extraordinary dielectric constants, respectively, and $\Delta=\epsilon_e-\epsilon_o$. Following Berreman's approach \cite{mccall_2015, berreman1972}, Maxwell equations are solved using the electric and magnetic field components parallel to the interfaces. For propagation along $y$, monochromatic waves take the form
\begin{equation} 
\begin{split}
    \vec{E}(x,y,t)=& \vec{E}\cdot e^{i(k_0(\kappa_xx+Ny)-\omega t)}\\
    \vec{H}(x,y,t)=& \vec{H}\cdot e^{i(k_0(\kappa_xx+Ny)-\omega t)},
\end{split}
\label{ef}
\end{equation}
with $k_0$ being the vacuum wavenumber.  In an isotropic material with dielectric constant $\epsilon$, the solutions reduce to TE and TM sets, with eigenvalues $\kappa_x=\kappa=\pm\sqrt{\epsilon - k_{y}^{2}}$, for forward (+ sign) and  backward (- sign) propagation. Then the field amplitudes in (\ref{ef}) can be described using a column vector $\vec{F}$ as
\begin{equation}
\vec{F}=\left[\begin{matrix}E_{y}\\H_{z}\\E_{z}\\H_{y}\end{matrix}\right],\quad\vec{F}_{TE}= \left[\begin{matrix}1\\\frac{\epsilon}{\kappa}\\0\\0\end{matrix}\right], \quad\vec{F}_{TM}=\left[\begin{matrix}0\\0\\1\\- \kappa\end{matrix}\right].
\end{equation}
\medskip
In anisotropic uniaxial materials the solutions are ordinary and  extraordinary waves. Here we use the corresponding analytic expression derived in \cite{mukherjee_topological_2018} with ordinary $\kappa_o$ and extraordinary $\kappa_e$ eigenvalues\\
\begin{equation}
\kappa_o=\pm\sqrt{\epsilon - k_{y}^{2}},
\end{equation}
\begin{equation}\kappa_e=\frac{-1}{\epsilon_{xx}} \left(\epsilon_{xy} k_{y} \pm \sqrt{\epsilon_{o}{\left (\epsilon_{xx} \epsilon_{e} + k_{y}^{2} \left(\epsilon_{zz} - \epsilon_{e} - \epsilon_{o}\right) \right )}}\right),
\end{equation}
with
\begin{equation}
\begin{split}
\vec{F}_o & =\left[\begin{matrix}\kappa_{o} \sin{\left (\phi \right )} \sin{\left (\theta \right )}\\\epsilon_{o} \sin{\left (\phi \right )} \sin{\left (\theta \right )}\\k_{y} \cos{\left (\theta \right )} - \kappa_{o} \sin{\left (\theta \right )} \cos{\left (\phi \right )}\\- \kappa_{o} \left(k_{y} \cos{\left (\theta \right )} - \kappa_{o} \sin{\left (\theta \right )} \cos{\left (\phi \right )}\right)\end{matrix}\right], \\
\vec{F}_e & =\left[\begin{matrix}- k_{y} \kappa_{e} \cos{\left (\theta \right )} + \kappa_{o}^{2} \sin{\left (\theta \right )} \cos{\left (\phi \right )}\\\epsilon_{o} \left(- k_{y} \cos{\left (\theta \right )} + \kappa_{e} \sin{\left (\theta \right )} \cos{\left (\phi \right )}\right)\\\epsilon_{o} \sin{\left (\phi \right )} \sin{\left (\theta \right )}\\- \epsilon_{o} \kappa_{e} \sin{\left (\phi \right )} \sin{\left (\theta \right )}\end{matrix}\right].
\end{split}
\end{equation}

Following Berreman's approach, a layer of material is accounted for by its field 4x4 matrix $\hat{F}$, composed by the four field vectors $\vec{F}$ representing the forward (+ superindex) and backward (- superindex) waves
\begin{equation}
\begin{split}
\hat{F}_{Iso} & =\left[\vec{F}_{TE}^{+}, \vec{F}_{TE}^{-}, \vec{F}_{TM}^{+}, \vec{F}_{TM}^{-}\right],\\ \hat{F}_{Ux} & =\left[\vec{F}_{e}^{+}, \vec{F}_{e}^{-}, \vec{F}_{o}^{+}, \vec{F}_{o}^{-}\right].
\end{split}
\end{equation}
The optical response of a multilayer system is described by the  characteristic matrix $\hat{A}$ calculated as
\begin{equation}
    \hat{A}=\hat{F}_c^{-1}\hat{M}\hat{F}_s,
\end{equation}
where $\hat{F}_c$ and $\hat{F}_s$ are the cladding and substrate field matrices respectively, and $\hat{M}$ is defined for a stack of N layers as
\begin{equation}
    \hat{M}=\hat{F}_1\hat{A}_1\hat{F}_1^{-1}\hat{F}_2\hat{A}_2\hat{F}_2^{-1}... \hat{F}_N\hat{A}_N\hat{F}_N1^{-1}.
\end{equation}
The diagonal matrix $\hat{A}_i$ is the phase matrix of layer $i$, which in an isotropic layer writes
\begin{equation}
    \hat{A}_i=\left[\begin{matrix}e^{i d_i k_{0} \kappa} & 0 & 0 & 0\\0 & e^{-i d_i k_{0} \kappa} & 0 & 0\\0 & 0 & e^{i d_i k_{0} \kappa} & 0\\0 & 0 & 0 & e^{-i d_i k_{0} \kappa}\end{matrix}\right],
\end{equation}
and in an uniaxial layer it writes
\begin{equation}
    \hat{A}_i=\left[\begin{matrix}e^{i d_i k_{0} \kappa_e^+} & 0 & 0 & 0\\0 & e^{i d_i k_{0} \kappa_e^-} & 0 & 0\\0 & 0 & e^{i d_i k_{0} \kappa_o} & 0\\0 & 0 & 0 & e^{-i d_i k_{0} \kappa_o}\end{matrix}\right],
\end{equation}
with $d_i$ being the thickness of the $i^{th}$ layer.\\

\indent For guided modes, the evanescent waves in the cladding are related to the evanescent waves of the substrate by the characteristic matrix as
\begin{equation}
\begin{split}
    A_{TE,s}^+\vec{F}_{TE}^+ + & A_{TM,s}^+\vec{F}_{TM}^+ =\\ 
     & \hat{M}(A_{TE,c}^-\vec{F}_{TE}^- + A_{TM,c}^-\vec{F}_{TM}^-),
\end{split}
\end{equation}
where $A_{TE,s}^+$ and $A_{TM,s}^+$ are the amplitudes of the forward TE and TM waves in the substrate and $A_{TE,c}^-$ and $A_{TM,c}^-$ the  amplitudes of the backward propagating TE and TM waves in the cladding. This results in a system of equations describing the eigenvalue problem, characterized by the system matrix.

\indent The calculation of $\hat{M}$ can be readily done numerically. However, it is instructive to derive the modal equation from the boundary conditions for a layer system comprising a uniaxial film and isotropic cladding and substrate, and write it in the form of a 8x8 matrix. At the interface between the substrate and the film, the boundary conditions write
\begin{equation}
\begin{split}
    A_{TE,s}^+\vec{F}_{TE}^+ & + A_{TM,s}^+\vec{F}_{TM}^+ = \\
    & A_{e}^+\vec{F}_{e}^+e^{i \frac{d}{2} k_{0} \kappa_e^+} + A_{e}^-\vec{F}_{e}^-e^{i \frac{d}{2} k_{0} \kappa_e^-} + \\ 
    & A_{o}^+\vec{F}_{o}^+e^{i \frac{d}{2} k_{0} \kappa_o} + A_{o}^-\vec{F}_{o}^-e^{-i \frac{d}{2} k_{0} \kappa_o},
\end{split}
\label{eqbcf}
\end{equation}
and at the interface between cladding and film they write
\begin{equation}
\begin{split}
    A_{TE,c}^-\vec{F}_{TE}^- & + A_{TM,c}^-\vec{F}_{TM}^- = \\
    & A_{e}^+\vec{F}_{e}^+e^{-i \frac{d}{2} k_{0} \kappa_e^+} + A_{e}^-\vec{F}_{e}^-e^{-i \frac{d}{2} k_{0} \kappa_e^-} + \\
    & A_{o}^+\vec{F}_{o}^+e^{-i \frac{d}{2} k_{0} \kappa_o} + A_{o}^-\vec{F}_{o}^-e^{+i \frac{d}{2} k_{0} \kappa_o},
\end{split}
\label{eqbcc}
\end{equation}
where $A_{e}^+$, $A_{e}^-$, $A_{o}^+$ and $A_{o}^-$ are the amplitudes of the four waves in the film.

Equations (\ref{eqbcf}) and (\ref{eqbcc}) can be written as a 8x8 matrix, the determinant of which yields the eigenvalue equation. In general, modes need all 8 amplitudes, and the whole matrix cannot be separated into smaller blocks. Thus, anisotropy couples the eight waves. However, under special conditions the matrix can be written as  4x4 blocks. Then the system matrix takes the form
\begin{equation}
\left[\begin{matrix} \hat{B}_1 \quad \hat{O}_1\\ \hat{O}_2   \quad \hat{B}_2\end{matrix}\right]\vec{A}_m=\vec{0}.
\label{bobo}
\end{equation}

Specifically, for $\phi=0^{\circ}$ ($x-y$ plane), the 8x8 matrix splits by polarization, corresponding to the case shown in Fig.~\ref{f3}(a). Under such conditions, all components in $\hat{O}_1$ and $\hat{O}_2$ vanish, even for asymmetric structures (Fig.~\ref{f5}(a)). Then, DPs exist in both situation as the matrix can be split in two blocks. For the $TE$ polarization one gets 
\begin{widetext}
\begin{equation} 
    \hat{B}_1\left[\begin{matrix} A_{TE,s}^+\\A_{e}^+\\A_{e}^-\\A_{TE,c}^-\end{matrix}\right]=\left[\begin{matrix}1 & \left(k_{y} \kappa^{+}_{e} \cos{\left (\theta \right )} - \kappa_{o}^{2} \sin{\left (\theta \right )}\right) e^{\frac{i d}{2} k_{0} \kappa^{+}_{e}} & \left(k_{y} \kappa^{-}_{e} \cos{\left (\theta \right )} - \kappa_{o}^{2} \sin{\left (\theta \right )}\right) e^{\frac{i d}{2} k_{0} \kappa^{-}_{e}} & 0\\- \frac{\epsilon_{s}}{\kappa_{s}} & \epsilon_{o} \left(k_{y} \cos{\left (\theta \right )} - \kappa^{+}_{e} \sin{\left (\theta \right )}\right) e^{\frac{i d}{2} k_{0} \kappa^{+}_{e}} & \epsilon_{o} \left(k_{y} \cos{\left (\theta \right )} - \kappa^{-}_{e} \sin{\left (\theta \right )}\right) e^{\frac{i d}{2} k_{0} \kappa^{-}_{e}} & 0\\0 & \left( k_{y} \kappa^{+}_{e} \cos{\left (\theta \right )} - \kappa_{o}^{2} \sin{\left (\theta \right )}\right) e^{- \frac{i d}{2} k_{0} \kappa^{+}_{e}} & \left( k_{y} \kappa^{-}_{e} \cos{\left (\theta \right )} - \kappa_{o}^{2} \sin{\left (\theta \right )}\right) e^{- \frac{i d}{2} k_{0} \kappa^{-}_{e}} & 1\\0 & \epsilon_{o} \left( k_{y} \cos{\left (\theta \right )} - \kappa^{+}_{e} \sin{\left (\theta \right )}\right) e^{- \frac{i d}{2} k_{0} \kappa^{+}_{e}} & \epsilon_{o} \left( k_{y} \cos{\left (\theta \right )} - \kappa^{-}_{e} \sin{\left (\theta \right )}\right) e^{- \frac{i d}{2} k_{0} \kappa^{-}_{e}} &  \frac{\epsilon_{c}}{\kappa_{c}}\end{matrix}\right]\left[\begin{matrix} A_{TE,s}^+\\A_{e}^+\\A_{e}^-\\A_{TE,c}^-\end{matrix}\right]=\vec{0},
\end{equation}
and for $TM$ polarization one gets
\begin{equation} 
    \hat{B}_2\left[\begin{matrix} A_{TM,s}^+\\A_{o}^+\\A_{o}^-\\A_{TM,c}^-\end{matrix}\right]=\left[\begin{matrix}1 & \left(- k_{y} \cos{\left (\theta \right )} + \kappa_{o} \sin{\left (\theta \right )}\right) e^{\frac{i d}{2} k_{0} \kappa_{o}} & - \left(k_{y} \cos{\left (\theta \right )} + \kappa_{o} \sin{\left (\theta \right )}\right) e^{- \frac{i d}{2} k_{0} \kappa_{o}} & 0\\\kappa_{s} & \kappa_{o} \left(k_{y} \cos{\left (\theta \right )} - \kappa_{o} \sin{\left (\theta \right )}\right) e^{\frac{i d}{2} k_{0} \kappa_{o}} & - \kappa_{o} \left(k_{y} \cos{\left (\theta \right )} + \kappa_{o} \sin{\left (\theta \right )}\right) e^{- \frac{i d}{2} k_{0} \kappa_{o}} & 0\\0 & \left(k_{y} \cos{\left (\theta \right )} - \kappa_{o} \sin{\left (\theta \right )}\right) e^{- \frac{i d}{2} k_{0} \kappa_{o}} & \left(k_{y} \cos{\left (\theta \right )} + \kappa_{o} \sin{\left (\theta \right )}\right) e^{\frac{i d}{2} k_{0} \kappa_{o}} & -1\\0 & \kappa_{o} \left(- k_{y} \cos{\left (\theta \right )} + \kappa_{o} \sin{\left (\theta \right )}\right) e^{- \frac{i d}{2} k_{0} \kappa_{o}} & \kappa_{o} \left(k_{y} \cos{\left (\theta \right )} + \kappa_{o} \sin{\left (\theta \right )}\right) e^{\frac{i d}{2} k_{0} \kappa_{o}} & \kappa_{c}\end{matrix}\right]\left[\begin{matrix} A_{TM,s}^+\\A_{o}^+\\A_{o}^-\\A_{TM,c}^-\end{matrix}\right]=\vec{0}.
\end{equation}
\\
Another example of matrix splitting in blocks occurs at $\theta=90^{\circ}$ ($y-z$ plane), where eigenmodes can be described as  odd and even modes. Algebraic combinations of the amplitudes $\vec{A}_m$ lead to the new base
\\
\begin{equation}
\begin{matrix}
   & & A_{TE}^{sum} =A_{TE,s}^+ + A_{TE,c}^-, 
   & A_{TE}^{sub.}=A_{TE,s}^+ - A_{TE,c}^-, 
   & A_{TM}^{sum}=A_{TM,s}^+ + A_{TM,c}^-,
   & A_{TM}^{sub.}=A_{TM,s}^+ - A_{TM,c}^-, \\
   & & A_{e}^{sum}=A_{e}^+ + A_{e}^-,
   & A_{e}^{sub.}=A_{e}^+ - A_{e}^-,
   & A_{o}^{sum}=A_{o}^+ + A_{o}^-,
   & A_{o}^{sub.}=A_{o}^+ - A_{o}^-.
\end{matrix}
\label{e1}
\end{equation}
Then, the diagonal blocks $\hat{B}_1$ and $\hat{B}_2$ in (\ref{bobo}) write

\begin{equation}
    \hat{B}_1\left[\begin{matrix} A_{TE}^{sum}\\A_{TM}^{sub.}\\A_{o}^{sum}\\A_{e}^{sub.}\end{matrix}\right]=\left[\begin{matrix}2 & 0 & - \kappa_{o} \sin{\left (\phi \right )} \cos{\left (\frac{d k_{0}}{2} \kappa_{o} \right )} & - \kappa_{o}^{2} \cos{\left (\phi \right )} \cos{\left (\frac{d k_{0}}{2} \kappa_{e} \right )}\\- \frac{\epsilon_{c}}{\kappa_{c}} - \frac{\epsilon_{s}}{\kappa_{s}} & 0 & - i \epsilon_{o} \sin{\left (\phi \right )} \sin{\left (\frac{d k_{0}}{2} \kappa_{o} \right )} & - i \epsilon_{o} \kappa_{e} \sin{\left (\frac{d k_{0}}{2} \kappa_{e} \right )} \cos{\left (\phi \right )}\\0 & 2 & \kappa_{o} \cos{\left (\phi \right )} \cos{\left (\frac{d k_{0}}{2} \kappa_{o} \right )} & - \epsilon_{o} \sin{\left (\phi \right )} \cos{\left (\frac{d k_{0}}{2} \kappa_{e} \right )}\\0 & \kappa_{c} + \kappa_{s} & - i \kappa_{o}^{2} \sin{\left (\frac{d k_{0}}{2} \kappa_{o} \right )} \cos{\left (\phi \right )} & i \epsilon_{o} \kappa_{e} \sin{\left (\phi \right )} \sin{\left (\frac{d k_{0}}{2} \kappa_{e} \right )}\end{matrix}\right]\left[\begin{matrix} A_{TE}^{sum}\\A_{TM}^{sub.}\\A_{o}^{sum}\\A_{e}^{sub.}\end{matrix}\right]=\vec{0},
\end{equation}
and
\begin{equation}
     \hat{B}_2\left[\begin{matrix} A_{TE}^{sub.}\\A_{TM}^{sum}\\A_{o}^{sub.}\\A_{e}^{sum}\end{matrix}\right]=\left[\begin{matrix}2 & 0 & - i \kappa_{o} \sin{\left (\phi \right )} \sin{\left (\frac{d k_{0}}{2} \kappa_{o} \right )} & - i \kappa_{o}^{2} \sin{\left (\frac{d k_{0}}{2} \kappa_{e} \right )} \cos{\left (\phi \right )}\\- \frac{\epsilon_{c}}{\kappa_{c}} - \frac{\epsilon_{s}}{\kappa_{s}} & 0 & - \epsilon_{o} \sin{\left (\phi \right )} \cos{\left (\frac{d k_{0}}{2} \kappa_{o} \right )} & - \epsilon_{o} \kappa_{e} \cos{\left (\phi \right )} \cos{\left (\frac{d k_{0}}{2} \kappa_{e} \right )}\\0 & 2 & i \kappa_{o} \sin{\left (\frac{d k_{0}}{2} \kappa_{o} \right )} \cos{\left (\phi \right )} & - i \epsilon_{o} \sin{\left (\phi \right )} \sin{\left (\frac{d k_{0}}{2} \kappa_{e} \right )}\\0 & \kappa_{c} + \kappa_{s} & - \kappa_{o}^{2} \cos{\left (\phi \right )} \cos{\left (\frac{d k_{0}}{2} \kappa_{o} \right )} & \epsilon_{o} \kappa_{e} \sin{\left (\phi \right )} \cos{\left (\frac{d k_{0}}{2} \kappa_{e} \right )}\end{matrix}\right]\left[\begin{matrix} A_{TE}^{sub.}\\A_{TM}^{sum}\\A_{o}^{sub.}\\A_{e}^{sum}\end{matrix}\right]=\vec{0}.
\end{equation}
\end{widetext}
In this case, blocks $\hat{O}_1$ and $\hat{O}_2$ in (\ref{bobo}) are
\begin{equation}
    \hat{O}_1=\hat{O}_2=\left[\begin{matrix}0 & 0 & 0 & 0\\\frac{\epsilon_{c}}{\kappa_{c}} - \frac{\epsilon_{s}}{\kappa_{s}} & 0 & 0 & 0\\0 & 0 & 0 & 0\\0 & - \kappa_{c} + \kappa_{s} & 0 & 0\end{matrix}\right].
    \label{oo}
\end{equation}
These two blocks can vanish only when $\epsilon_s=\epsilon_c$ and $\kappa_s=\kappa_c$, i.e., when the substrate and cladding are identical and the amplitudes in (\ref{e1}) describe even and odd modes. If the waveguide is asymmetrical, the matrix (\ref{oo}) does not vanish and thus the system matrix (\ref{bobo}) does not split in blocks and, therefore, DPs cannot exist. This is the case shown in Fig.~\ref{f5}(c), where instead of a DP, a gap is opened in the dispersion diagram.  

Finally, another situation of interest occurs for $\phi=90^{\circ}$. This corresponds to the DP shown in Figures \ref{f2}(a) and \ref{f2}(c) in the main text. Here one can use again the definitions in (\ref{e1}) to find the same blocks $\hat{O}_1$ and $\hat{O}_2$ as in (\ref{oo}), while the diagonal blocks $\hat{B}_1$ and $\hat{B}_2$ in (\ref{bobo}) write
\begin{widetext}
\begin{equation}
    \hat{B}_1\left[\begin{matrix} A_{TE}^{sub.}\\A_{TM}^{sum}\\A_{o}^{sum}\\A_{e}^{sub.}\end{matrix}\right]=\left[\begin{matrix}2 & 0 & - i \kappa_{o} \sin{\left (\theta \right )} \sin{\left (\frac{d k_{0}}{2} \kappa_{o} \right )} & i k_{y} \kappa_{e} \sin{\left (\frac{d k_{0}}{2} \kappa_{e} \right )} \cos{\left (\theta \right )}\\- \frac{\epsilon_{c}}{\kappa_{c}} - \frac{\epsilon_{s}}{\kappa_{s}} & 0 & - \epsilon_{o} \sin{\left (\theta \right )} \cos{\left (\frac{d k_{0}}{2} \kappa_{o} \right )} & \epsilon_{o} k_{y} \cos{\left (\theta \right )} \cos{\left (\frac{d k_{0}}{2} \kappa_{e} \right )}\\0 & 2 & - k_{y} \cos{\left (\theta \right )} \cos{\left (\frac{d k_{0}}{2} \kappa_{o} \right )} & - \epsilon_{o} \sin{\left (\theta \right )} \cos{\left (\frac{d k_{0}}{2} \kappa_{e} \right )}\\0 & \kappa_{c} + \kappa_{s} & i k_{y} \kappa_{o} \sin{\left (\frac{d k_{0}}{2} \kappa_{o} \right )} \cos{\left (\theta \right )} & i \epsilon_{o} \kappa_{e} \sin{\left (\theta \right )} \sin{\left (\frac{d k_{0}}{2} \kappa_{e} \right )}\end{matrix}\right]\left[\begin{matrix} A_{TE}^{sub.}\\A_{TM}^{sum}\\A_{o}^{sum}\\A_{e}^{sub.}\end{matrix}\right]=\vec{0},
\end{equation}
and
\begin{equation}
    \hat{B}_2\left[\begin{matrix} A_{TE}^{sum}\\A_{TM}^{sub.}\\A_{o}^{sub.}\\A_{e}^{sum}\end{matrix}\right]=\left[\begin{matrix}2 & 0 & - \kappa_{o} \sin{\left (\theta \right )} \cos{\left (\frac{d k_{0}}{2} \kappa_{o} \right )} & k_{y} \kappa_{e} \cos{\left (\theta \right )} \cos{\left (\frac{d k_{0}}{2} \kappa_{e} \right )}\\- \frac{\epsilon_{c}}{\kappa_{c}} - \frac{\epsilon_{s}}{\kappa_{s}} & 0 & - i \epsilon_{o} \sin{\left (\theta \right )} \sin{\left (\frac{d k_{0}}{2} \kappa_{o} \right )} & i \epsilon_{o} k_{y} \sin{\left (\frac{d k_{0}}{2} \kappa_{e} \right )} \cos{\left (\theta \right )}\\0 & 2 & - i k_{y} \sin{\left (\frac{d k_{0}}{2} \kappa_{o} \right )} \cos{\left (\theta \right )} & - i \epsilon_{o} \sin{\left (\theta \right )} \sin{\left (\frac{d k_{0}}{2} \kappa_{e} \right )}\\0 & \kappa_{c} + \kappa_{s} & k_{y} \kappa_{o} \cos{\left (\theta \right )} \cos{\left (\frac{d k_{0}}{2} \kappa_{o} \right )} & \epsilon_{o} \kappa_{e} \sin{\left (\theta \right )} \cos{\left (\frac{d k_{0}}{2} \kappa_{e} \right )}\end{matrix}\right]\left[\begin{matrix} A_{TE}^{sum}\\A_{TM}^{sub.}\\A_{o}^{sub.}\\A_{e}^{sum}\end{matrix}\right]=\vec{0}.
\end{equation}
\end{widetext}
In this case the polar orientation of the optical axis can have an arbitrary value, which may break the anisotropy symmetry with respect to the $y-z$ plane \cite{mukherjee_topological_2018}. However, importantly, eigenmodes can also be expressed as even and odd modes, and DPs can exist provided the structure is symmetric in refractive index.

\end{document}